\documentclass[aps,pra,twocolumn,amsmath,amssymb,superscriptaddress,showpacs]{revtex4-1}

\usepackage{amsfonts}
\usepackage{amsmath}
\usepackage{amssymb}
\usepackage{amsthm}
\usepackage{fontenc}
\usepackage{graphicx}
\usepackage{xcolor}
\usepackage{textcomp}
\usepackage{epstopdf}
\usepackage{braket}
\usepackage{mathtools}
\usepackage{amsmath}
\usepackage{dcolumn}
\usepackage{multirow}
\usepackage{units}

\begin{document}

\title{Robustness of type-II Dirac cones in  biphenylene: from Nanoribbons to novel symmetric bilayer stackings}

\author{L. L. Lage}
\affiliation{Instituto de F\'isica, Universidade Federal Fluminense, Niter\'oi, Av. Litor\^{a}nea sn 24210-340, RJ-Brazil}
\author{O. Arroyo-Gasc\'on}
\affiliation{Instituto de Ciencia de Materiales de Madrid, Consejo Superior de Investigaciones Cient\'{\i}ficas, C/ Sor Juana In\'es de la Cruz 3, 28049 Madrid, Spain.}
\affiliation{GISC, Departamento de F\'{\i}sica de Materiales, Facultad de Ciencias Físicas, Universidad Complutense de Madrid, E-28040 Madrid, Spain.} 
\author{L. Chico}
\affiliation{GISC, Departamento de F\'{\i}sica de Materiales, Facultad de Ciencias Físicas, Universidad Complutense de Madrid, E-28040 Madrid, Spain.}
\author{A. Latgé}
\affiliation{Instituto de F\'isica, Universidade Federal Fluminense, Niter\'oi, Av. Litor\^{a}nea sn 24210-340, RJ-Brazil}

\date{\today}

\begin{abstract}
The electronic properties of one- and two-dimensional biphenylene-based systems, such as nanoribbons and bilayers, are studied within a unified approach. Besides the bilayer with direct (AA) stacking, we have found two additional symmetric stackings for bilayer biphenylene that we denote by AB, by analogy with bilayer graphene, and AX, which can be derived by a small translation (slip) from the AA bilayer,  with distinct electronic band structures.
We combine first-principles calculations with a tight-binding model to provide a realistic effective description of these biphenylene materials.  Our approach provides a global framework to analyze the robustness of the characteristic type-II Dirac cones present in all the bilayers studied, capturing 
the variations caused by different stackings. Additionally, we find that a pseudogap opens in the Dirac cone for certain nanoribbons, depending on its width. We relate this effect to the symmetries of the system.
\end{abstract}

\maketitle

%%%MAIN TEXT%%%%

\section{Introduction}\label{Intro}

Carbon-based 2D crystals, particularly those structured with hexagonal (benzene) rings, have garnered significant interest beyond graphene systems. These new allotropes include monolayer structures such as those synthesized in recent studies \cite{hu2022synthesis,LI2023100140}, bilayer structures
\cite{D1CP05491K,graphyne2023}, and graphdiyne \cite{Zheng2023}, which is characterized by the insertion of sp acetylenic bonds within the carbon lattice. Other carbon materials without hexagonal carbon rings have been theoretically proposed, such as pentaheptite \cite{Crespi1996}, completely composed of pentagons and heptagons; or a semiconducting planar sheet formed by 4- and 8-atom carbon rings \cite{nisar2012semiconducting}, and even a buckled 2D material made of distorted pentagons, known as pentagraphene \cite{Tang2014,Zhang2015}. These two latter examples have the interest of being semiconductor planar forms of carbon which could complement graphene. 
Additionally, structures combining hexagonal and other n-carbon rings have been also proposed \cite{Xu2014,bandyopadhyay2020topology}. Among them, biphenylene stands out due to its recent experimental synthesis via the dehydrofluorination fusion of benzenoid polyphenylene chains \cite{fan2021biphenylene}. This planar $sp^2$ carbon network structure was proposed long ago \cite{Balaban1968}. It exhibits an intricate geometry,  comprising 4-, 6- and 8-folded rings.  Electronic stability of biphenylene systems was also studied before its synthesis by means of first-principles calculations, including ribbons and tubes of different widths and morphologies \cite{hudspeth2010electronic}.

After its experimental discovery, many works have explored the physical and chemical properties of biphenylene \cite{luo2021first, ke2022biphenylene, demirci2022hydrogenated,Son2022,Niu2023}. 
Density functional theory (DFT) calculations identified a type-II Dirac cone with metallic character which could be useful for valleytronics, %related 
due to the existence of two bands with the same sign of the carrier velocity 
\cite{Penfei2021,alcon2022unveiling,bafekry2022biphenylene}. However, a band gap can be achieved by applying strain or doping the lattice \cite{hou2023opening,lee2021band}. Another way of producing a band gap is by doping with fluorine atoms; this technique can be also applied to tune the Dirac cone and change its character \cite{Mo2024}. Theoretical results predict a semiconductor behavior for the armchair nanoribbons for small sizes ($<2 $ nm), and metallic for both, zigzag and larger armchair ribbons  \cite{hudspeth2010electronic,fan2021biphenylene}. Regarding bilayer systems, the AA stacking %in 2D biphenylene 
has been modeled within a DFT approach, predicting a stable configuration with enhanced elastic characteristics compared to its  monolayer counterpart \cite{Chowdhury2022}. Further studies reveal that biphenylene may have  properties of practical interest,  such as being an anti-corrosion coating material with exceptional  oxygen atom adsorption and reasonable hydrophobicity \cite{keke2022}, or thermoelectric applications in the 2D system \cite{D3CP03088A} and in nanoribbons \cite{xie2023intrinsic, farzadian2022theoretical}. By using biphenylene nanoribbons as the building block, new porous 3D metallic carbon structure was also reported \cite{Sun2022}. More recently, the topological properties of biphenylene were studied by means of a simple tight-binding model, which allowed to verify topological phase transitions 
and explore higher-order topological properties of this material 
\cite{Wakabayashi2024}. 

Motivated by these works, we 
propose two novel symmetric bilayer configurations, obtained by lattice displacements of one of the layers and with minimal translational unit cells. To the best of our knowledge, these bilayer stackings, that we denominate here by AB, by analogy with bilayer graphene, and AX, which is found by displacing one layer along the $x$ direction starting from the direct stacking AA, have not been described yet.   We also explore biphenylene nanoribbons, providing a tight-binding parametrization valid for these bilayers with symmetric stackings as well as for the nanoribbons. We analyze the changes in the Dirac cones, showing the modifications produced by the interlayer interaction and by the size and symmetry effects related to the nanoribbons, such as an anticrossing that results in a gap opening in the Dirac cone. 
 Our main findings are the following:
(i) We propose two novel symmetric stackings for bilayer biphenylene, AB and AX, with comparable total energies with respect to the AA case;  
(ii) A type-II Dirac cone is reproduced by adopting our tight-binding parametrization. We have verified that the Dirac cone is robust and persists in bilayer biphenylene AA, AB and AX stackings; (iii) In armchair nanoribbons with an odd number of hexagons across their width, the Dirac crossing is preserved; for even widths, a gap opens in the cone. We relate this even-odd effect to the mirror symmetry of the wavefunction with respect to its longitudinal axis; 
(iv) A different number of in-gap topological edge states for wide armchair nanoribbons are obtained with our model compared to simpler parametrizations which is related to the inclusion of hopping energies between atoms up to 3 \AA \ apart.

We consider that our proposal of novel stackings can stimulate the experimental search of these bilayers and the study of their physical properties. Additionally, we expect that our tight-binding model will be of interest for further explorations of complex biphenylene structures with hybrid geometric structures.   

\section{Systems and methods }

In order to describe all the biphenylene structures, we start by performing a DFT relaxation of the geometries. With the relaxed coordinates, we perform a general tight-binding parametrization intended to be valid for all structures, i.e., 
applicable to the monolayer and also to nanoribbons and bilayers.  

\subsection{Geometry of monolayer biphenylene}

Differently from graphene, monolayer biphenylene belongs to the $Pmmm$ space group. It is described by a rectangular unit cell of orthogonal lattice vectors $\vec{a}_1$ and  $\vec{a}_2$, of unequal lengths $a_1\neq a_2$, with a six-atom basis, as depicted in Fig. \ref{FIG1} (a). Note that the hexagon is not regular, and neighboring atoms are at two different distances, $d_1$ and $d_2$. 
Repeating this pattern, octagons and 4-atom rings %show up
are distinguished in Fig. \ref{FIG1} (a), so four different distances between atoms are required. The first Brillouin zone, also rectangular, is depicted therein, with  the four high-symmetry points labeled, namely,  $\Gamma=(0,0)$, 
$X=(\frac{\pi}{{a}_1},0)$, $Y=(0,\frac{\pi}{{a}_2})$, and 
$S=(\frac{\pi}{{a}_1},\frac{\pi}{{a}_2})$.

\subsection{Geometries of bilayers with symmetric stackings}

\begin{figure}[!h]
    \centering
    \includegraphics[width=8.5cm]{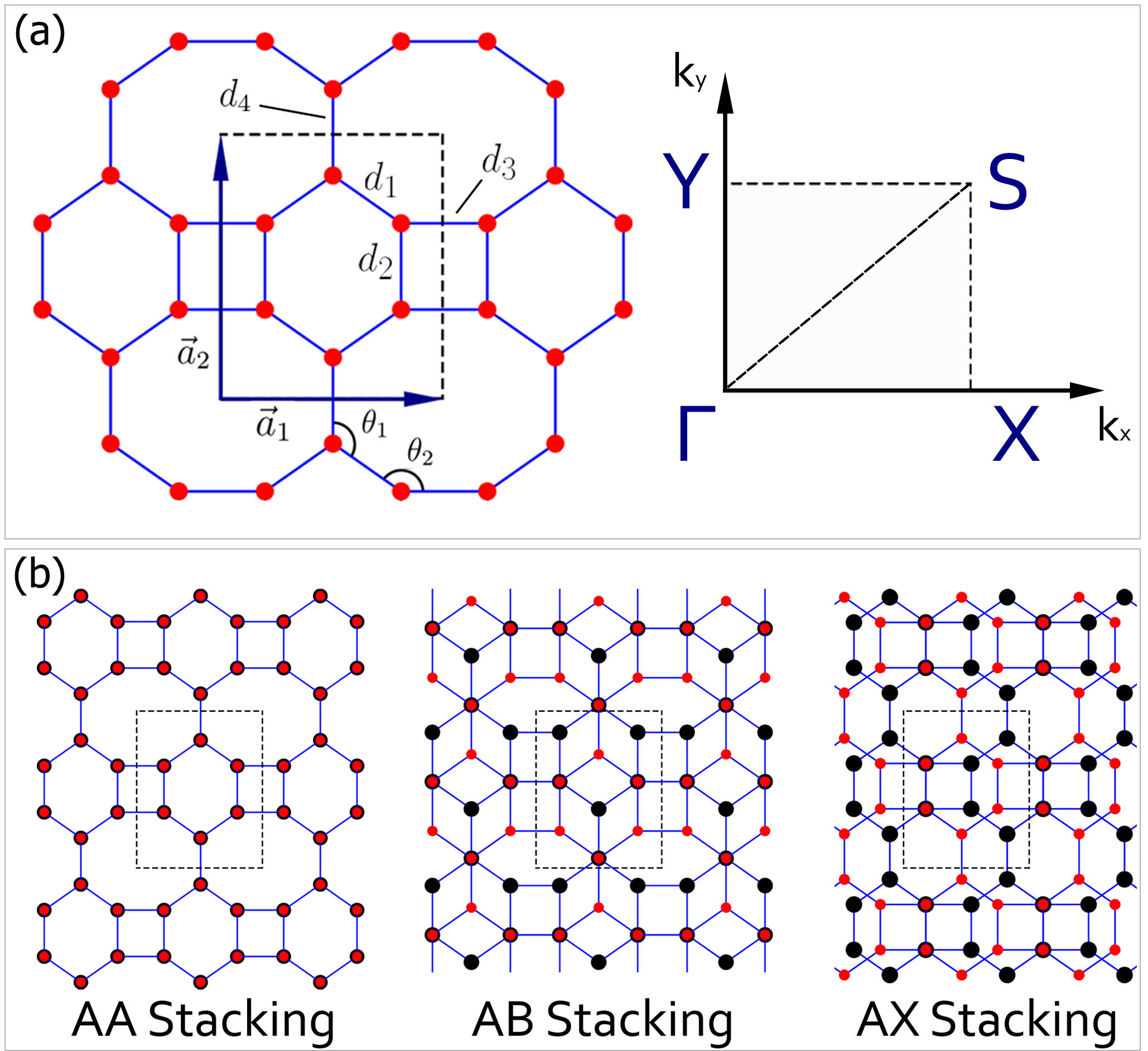}
\caption{(a) Left panel: 2D biphenylene lattice with primitive vectors $\vec{a}_1$ and $\vec{a}_2$ along the $x$ and $y$ direction, respectively.
Right panel: first Brillouin zone of biphenylene with special symmetry lines and points. (b) Schematic depiction of the three bilayer stackings studied in this work.}   \label{FIG1}
\end{figure}

The geometric stackings for bilayer biphenylene defined as  AA, AB, and AX are depicted in Fig. \ref{FIG1}(b), belonging to space groups, $Pmmm$, in the AA case, and $P2/m$ for AB and AX. Bilayer AA is the trivial, direct stacking, obtained by placing two biphenylene layers one on top of the other, studied in a previous work \cite{Chowdhury2022}; we also model this bilayer for completeness. 
We obtain the AB stacking from an AA bilayer, displacing one of the layers a distance $d_2$ along the $\vec{a}_2$ direction; see central panel of Fig.  \ref{FIG1}(b). We choose this denomination (AB)  because one atom belonging to a hexagon of the upper layer is at the center of the hexagon of the bottom layer, resembling thus the AB stacking of bilayer graphene. Finally, we derive the AX bilayer by starting from the AA configuration and displacing one layer by a distance 
$d_3$ along the $\vec{a}_1$ direction; it resembles the so-called slip stacking visible in graphene moiré patterns, as it can be seen in the right panel of Fig. \ref{FIG1}(b) \cite{Li2010,eric2011}. 

The novel proposed stackings AB and AX have similar total energies to the stable AA stacking within our DFT calculations, as detailed in the supplementary material. In all these cases the unit cell has 12 atoms, twice than that of the monolayer; the ideal (unrelaxed) lattice vectors are initially chosen to be the same as for the monolayer, and change after relaxation, as expected. The interlayer distance is close to that of graphene, and we obtained slightly different values after relaxation, as it will be discussed later.

\subsection{DFT calculations}

The SIESTA first-principles code \cite{siesta1,siesta2} is employed to perform electronic structure calculations, employing the generalized gradient approximation (GGA) and the Perdew–Burke-Ernzerhof (PBE) exchange-correlation functional \cite{GGA}. This is our functional of choice for the monolayer system. For the bilayer geometries, additional van der Waals functionals following Dion et al. \cite{VDWDion} with different improvements, namely DRSLL (equivalent to vdW-DF) \cite{DF1} and LMKLL (equivalent to vdW-DF2) \cite{DF2} flavors, were used. For all calculations, a double-$\zeta$ singly-polarized basis set was employed.
The reciprocal space was mapped by means of a $8 \times 8 \times 1$ Monkhorst-Pack grid for all systems. All structures were relaxed until the forces were below 0.01 eV/Å.

\subsection{Tight-binding approach}

Our goal with the tight-binding approach is to provide a unified description of all biphenylene structures.
Since we focus on the bands around the Fermi energy, a single $p_z$ orbital tight-binding (TB) Hamiltonian is used to describe the bilayer and monolayer systems, given by

\begin{gather}
H=\sum_{i,a}\varepsilon_{i}^{a}c_{i}^{\dagger a}c_{i}^{a}+\sum_{\substack { i,j\\ {a} }}
t_{ij}^{a}c_{i}^{\dagger a}c_{j}^{a}+\sum_{\substack { i,j \\ {a \neq b} }} t_{ij}^{\perp a b}c_{i}^{\dagger a}c_{j}^{b}+h.c.,\label{Hamito}
\end{gather}

\noindent where $\varepsilon_i^{a}$ is the onsite energy for each atom located at site $i$ in layer $a$, and the operator $c^{a\dagger}_{i}$ ($c^{a}_{i}$) creates (annihilates) an electron on site $i$ and layer $a$. The second term describes the intralayer couplings,
$t^{a}_{ij}$ being the corresponding hopping energies within layer $a$. Obviously for monolayers $a=1$, and for bilayers we consider two values $a(b)=1,2$, as  well as the interlayer interactions, given by the last term ($a\neq b$) and denoted as $t_{ij}^{\perp a b}$. They depend on the stacking configuration between top and bottom biphenylene layers. 

To find a suitable hopping parametrization we have considered an intralayer hopping energy described by a decaying exponential function \cite{eric2011}, 
\begin{equation}
    t^{a}_{ij}=t_1e^{-\beta\big(\frac{r_{ij}}{d_1}-1\big)} \ ,
\end{equation}
with $r_{ij}$ being the distance between $i,j$ lattice sites, $t_1$ the hopping related to the  
first %site 
nearest-neighbor distance $d_1$, and $\beta$ a fitting parameter that controls the range of the interaction. 
As the ratio $r_{ij}/d_1$ is always larger than one beyond the first nearest-neighbors, small $\beta$ values allows to increase the number of neighbors with non-negligible hoppings in the description.

For the interlayer connection we have also considered a decaying exponential function for the hopping energies given by
\begin{equation}
t_{ij}^{\perp a b}=t_0e^{-\alpha(r_{ij}-d_{\perp})} \ ,
\end{equation}
with $d_{\perp}$ being the interlayer distance, $t_0$ is the direct stacking hopping value, i.e., when the atoms are exactly one above the other, and $\alpha$ 
modulates the strength of the interlayer hopping with increasing distance.

\section{Two-dimensional biphenylene systems}

\subsection{Monolayer biphenylene}

The relaxed geometries obtained by DFT (PBE-GGA) for the monolayer are the following: the lengths of the lattice vectors are 
$a_1=3.82$ %15
\AA\ and $a_2=4.54$ %4
\AA , 
with octagon angles $\theta_1=125º$ and $\theta_2=145º$.   
The four distances between atoms with primary covalent bonds, defined in Fig. \ref{FIG1} (a), are given by  $d_1$=1.42 Å, $d_2$=1.44 Å, $d_3$=1.50 Å, and $d_4$=1.47 Å. The basis vectors can be written as $\vec{a}_1=(2 d_1 \sin(\pi - \theta_2) +d_3,0)$
% \sin[\frac{11 \pi}{36}] +d_3
and $\vec{a}_2=2(0, d_1 \cos(\pi - \theta_2)+ d_2)$. 

\begin{figure}[!h]
    \centering
    \includegraphics[width=8.5cm]{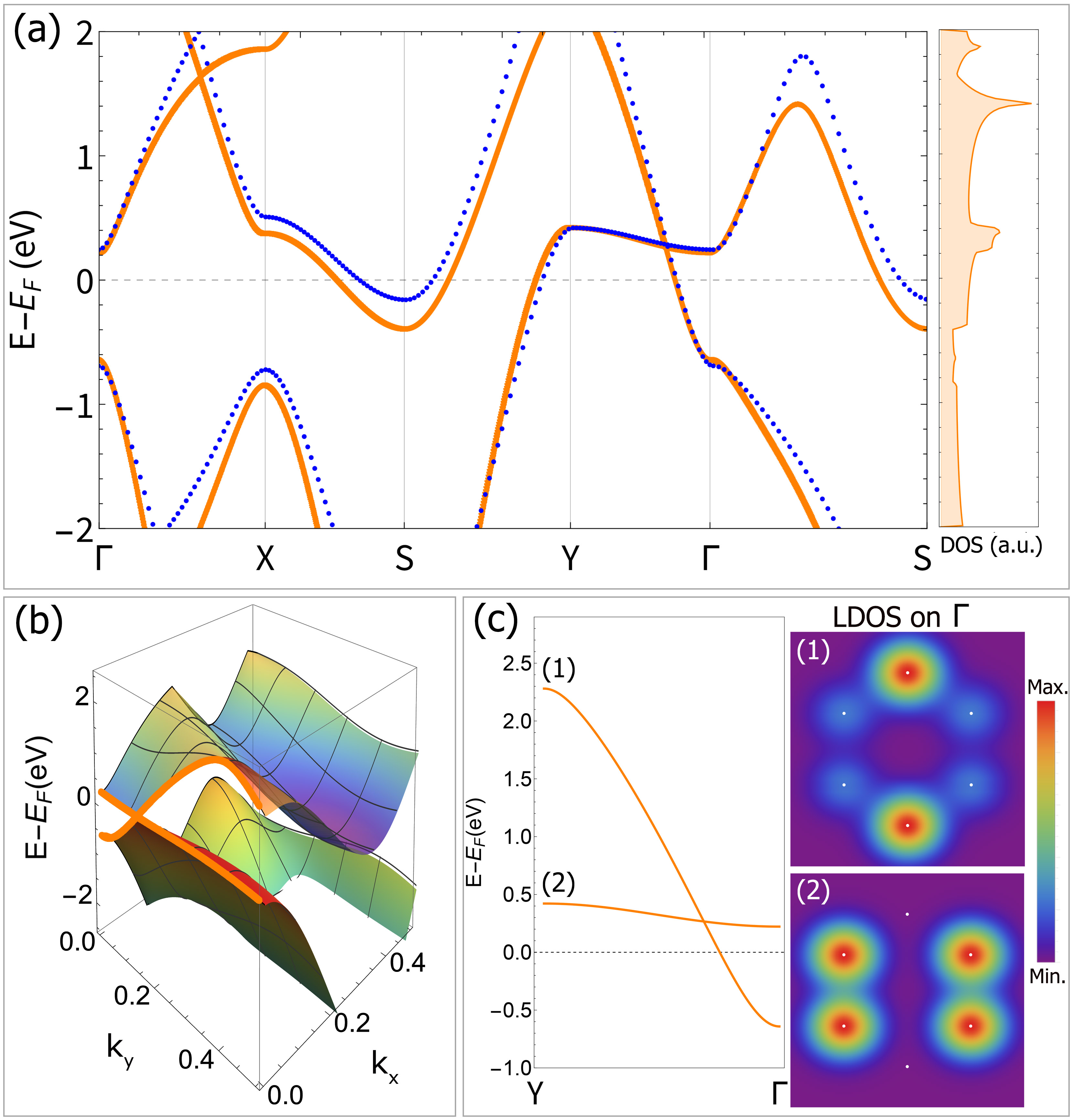}%v2fig2
   \caption{(a) Biphenylene electronic structure: DFT calculations (blue dotted curves) and TB results (orange bands) considering onsite energy $\varepsilon_1=-2.2$ eV and $\varepsilon_2=-1.85$ eV, for the sites with configurations (1) and (2) in Fig.\ref{fig2}-(c), respectively, $t_1=-3.3$ eV and $\beta=2.2$. The  DOS corresponding to the TB bands is plotted at the right side of the panel. (b) 2D tight-binding biphenylene energy bands, showing the type-II Dirac cone in orange. (c) Zoom of the band structure along the $Y-\Gamma$ high-symmetry line focusing on the type-II Dirac cone. At the right of the panel, the LDOS for the two bands forming the Dirac cone is depicted. The position of the atoms inside the unit cell are marked with white dots. Red and purple corresponds to the maximum and minimum probability density, respectively.}
   \label{fig2}
\end{figure}

We present in Fig. \ref{fig2}(a) a comparison between DFT PBE-GGA (dotted blue curves) and the fitted tight-binding  (full orange lines) band calculations. The agreement is very good for the fitting parameters shown in Table \ref{table1}, specially in the energy range close to the Fermi level. The density of states (DOS) obtained from the TB approach is displayed at the right of the electronic bands. The peculiar type-II Dirac cone appearing in the middle of the $Y-\Gamma$ path can be clearly seen in Fig. \ref{fig2}(a), and it is correctly described by the TB model. A 2D plot of the band structure is depicted in Fig. \ref{fig2}(b) where the type-II cone is highlighted in orange.

A zoom of the bands along $Y-\Gamma$ constituting the type-II Dirac cone, labeled (1) and (2), is presented in Fig. \ref{fig2}(c). The local density of states (LDOS) of the respective bands, calculated at the Y-point, is depicted at the side of the electronic bands. 
Note that the LDOS of the two bands  are localized in different regions of the unit cell, having maxima at different atoms. The two atoms with maximum LDOS in the top panel of Fig. \ref{fig2}(c), corresponding to band (1) are related by a mirror symmetry of the system. Likewise, the four atoms with maximum LDOS of the bottom panel, corresponding to the band (2), are also related by mirror symmetries of the 2D crystal. However, there is no symmetry operation that relates these two sets of atoms \cite{padilha}. Therefore, these states do not interact, and the bands cross, as seen in the left panel of Fig. \ref{fig2}(c). At the crossing energy (not shown here), all sites have a nonzero density. 
Since these two sets of atoms are not related by symmetry, a small variation in their respective on-site energies can be included to fit with our DFT results; $\varepsilon_1=-2.2$ eV and $\varepsilon_2=-1.85$eV for the respective sets.  It is important to emphasize that the use of a tight-binding parametrization with an exponential decay which includes hoppings between atoms up to 3 \AA \  is crucial in our model to obtain the predicted type-II Dirac cone in biphenylene and similar materials, in agreement with DFT calculations. 

\begin{table}[h]
\small
\caption{\label{tab:table1}%
Lattice parameters for the monolayer obtained from DFT (PBE-GGA) relaxation and fitted parameters for the TB calculations.
}\label{table1}
  \begin{tabular*}{0.48\textwidth}{@{\extracolsep{\fill}}ll}
    \hline
    Lattice Parameters & 2D TB Parameters \\
    \hline
$d_1=1.42$ Å & $t_1=-3.3$ eV \\
$d_2=1.44$ Å & $t_0=-0.33$ eV \\
$d_3=1.50$ Å & $\alpha=5.0$  \\
$d_4=1.47$ Å & $\beta=2.2$ \\
    \hline
  \end{tabular*}
\end{table}

\begin{figure}[!h]
    \centering
    \includegraphics[width=8.5cm]{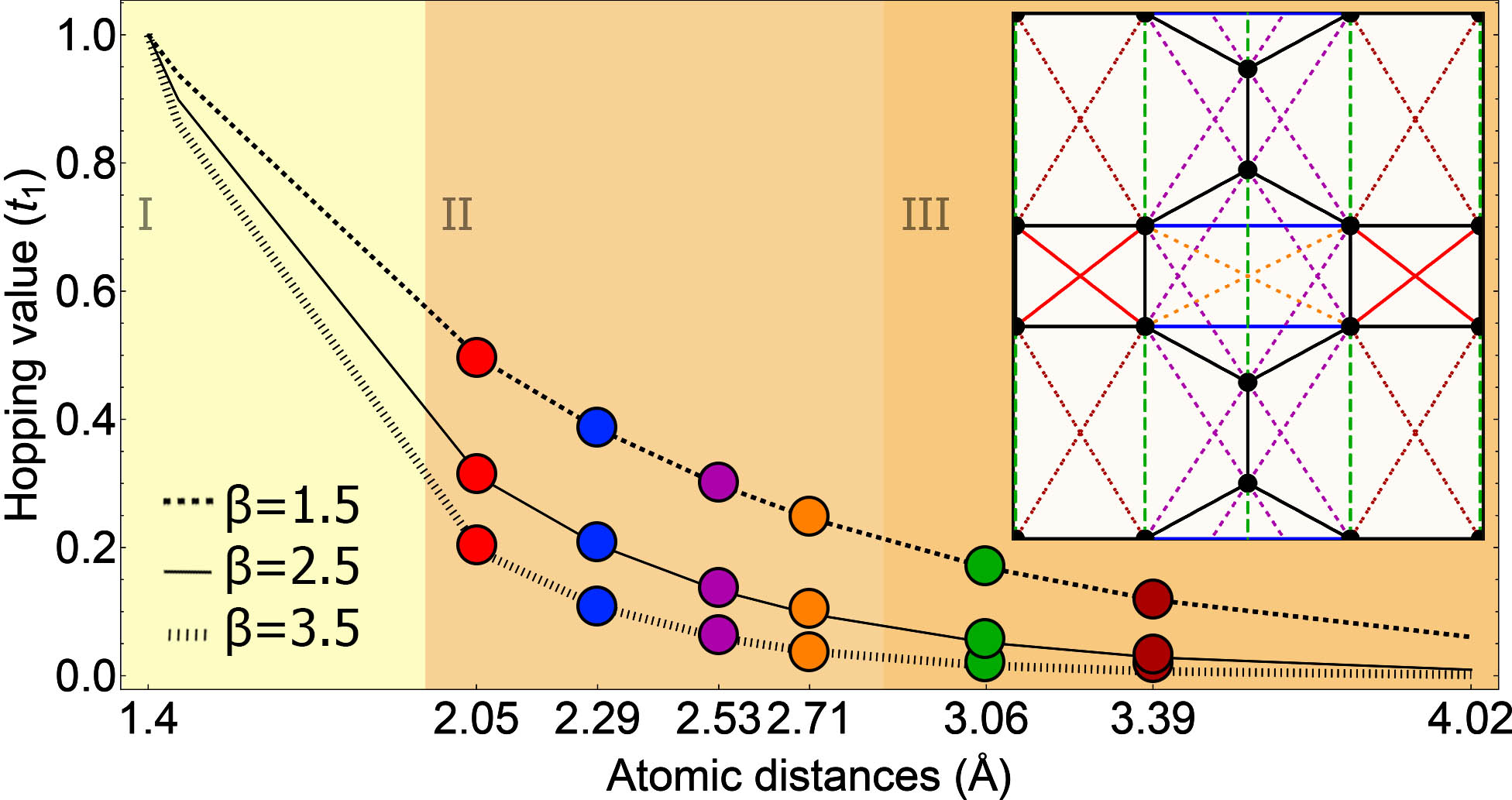}%expfig1
\caption{Illustration of in-plane hopping energies as a function of the atomic lattice distances for different values of the  decay parameter $\beta$. Colored circles illustrate hoppings at different lattice distances,  matching the color code employed  in the schematic lattice geometry shown in the inset. The 
 distances/hoppings are grouped by shaded regions, I, II, and III, respectively.}   \label{FIG3}
\end{figure}

The overall picture of the hopping decay parametrization with the neighbor distances is shown in Fig. \ref{FIG3}, with the hopping scheme in the inset, where each group of neighbors is highlighted with colored regions, I, II, and III. Region I corresponds to hoppings between atoms with primary bonds (black solid lines), region II is for intermediate distances (colored solid lines) and region III corresponds to long-range hoppings (colored dashed lines).

The slope of the bands is determined by the $\beta$ parameter, and altering it affects carrier velocities.
This adjustment can be tuned to achieve the desired characteristics of the bands. 
Additionally, it is essential to take into account a sufficient number of neighbors. 
 To achieve the desired type-II Dirac cone we must include hoppings up to region III, which is equivalent to consider distances greater than those marked by green dashed lines ($>$ 3 \AA) in the schematic inset. 
Not considering enough hopping parameters results in a completely flat energy band along the $Y-\Gamma$ direction in one of the branches of the Dirac cone, changing its character to type-III \cite{Wakabayashi2024, Mo2024}. Therefore, a careful choice of both the $\beta$ parameter, and hence the range of neighbors included, is critical to obtain the desired band features. We consider that this parametrization approach will also be helpful in other systems, where intricate symmetries and numerous hopping distances may hinder the derivation of optimal physical parameters for theories based in a TB model theory of the structure.    

\subsection{Bilayer biphenylene: AA, AB and AX stackings}

For bilayer biphenylene systems, we have included the van der Waals interaction in the DFT calculations by means of a vdW-DF2 functional. As in the case of the monolayer, the lattice parameters included in TB calculations were obtained from previous DFT relaxations. We find that the intralayer and the internal angles are the same as for the monolayer.

\begin{figure*}[t!]
    \centering
    \includegraphics[width=16cm, height=7cm]
    {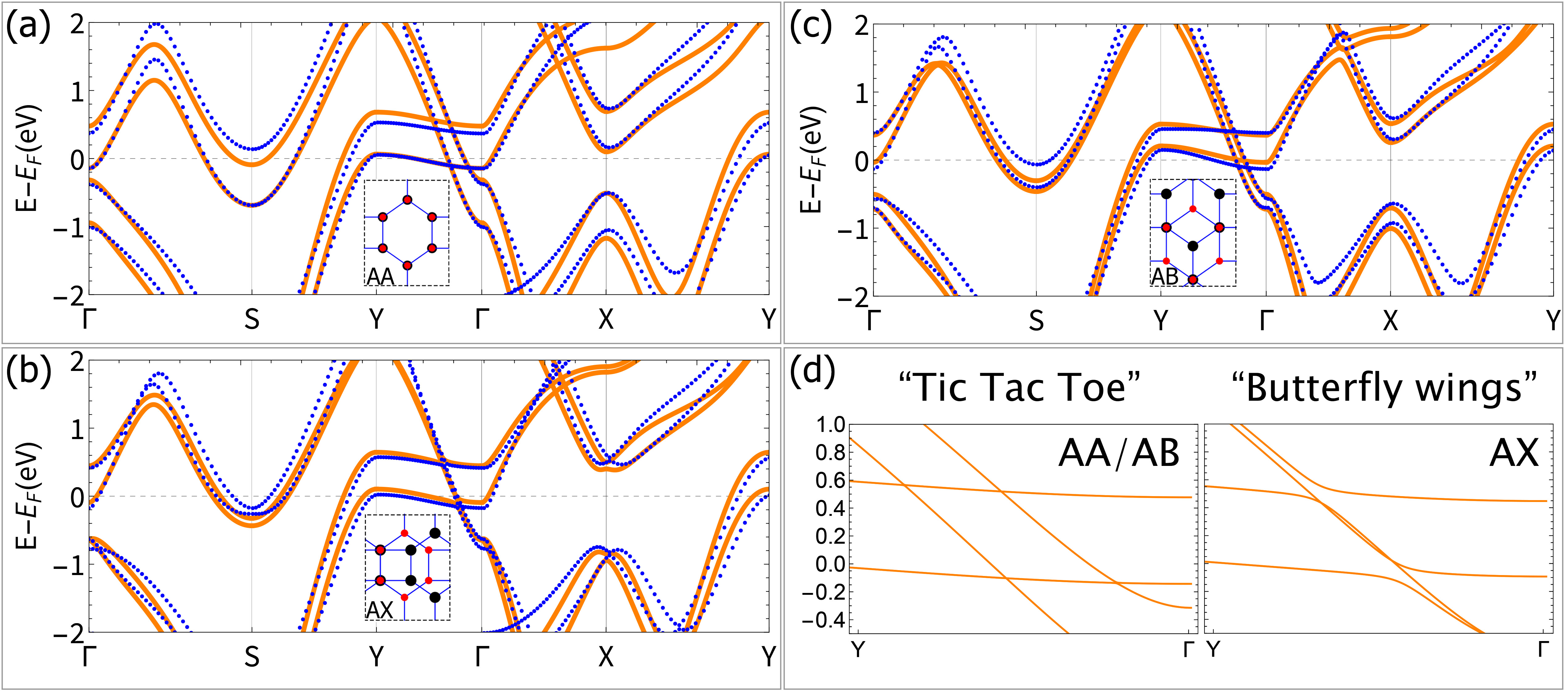}
     \caption{Bilayer biphenylene band structures for the three stackings considered: (a) AA, (b) AB, and (c) AX stackings. The corresponding unit cells are shown in the insets, marked with dashed lines. DFT bands are shown with blue dots, and TB as orange curves; the parameters chosen are $\alpha=5.0$, $\beta=2.2$ $t_{0}=-0.33$ eV, $t_1=-3.3$ eV, $\varepsilon_1=-2.2$ eV and $\varepsilon_2=-1.85$ eV.} \label{FIG4}
\end{figure*}

Table \ref{tablebilayer} displays the interlayer $d_{\perp}$ distances in each stacking. The unit vectors in the bilayer predicted by DFT are $a_1=3.84 $ Å and $a_2=4.54 $ Å. 
Previous DFT calculations \cite{Chowdhury2022} predicted a biphenylene interlayer distance of 3.36 Å in AA stacking, a value notably close to that of bilayer graphene (3.42 Å). 
However, our calculations yield a larger distance for the AA stacking. The values for the three studied stackings reveal a small variation for the interlayer distances, as summarized in Table \ref{tablebilayer}. In our calculation, the AB stacking has the smallest interlayer spacing. 
All bilayer geometries were found to be stable and had comparable total energies. 
The smallest interlayer distances correspond to the vdW-DF2 functional. In the supplementary material we show a comparison of the energy bands computed using different DFT functionals, in particular, GGA-PBE, vdW-DF, and vdW-DF2. The used TB parameters are $\alpha=5.0$ and $t_{0}=-0.33$ eV for the interlayer coupling; the same intralayer parameters used in the case of the monolayer are adopted.

 In Fig. \ref{FIG4} (a)-(c) we compare the DFT and TB band structures of the three stacking configurations, AA, AX and AB, respectively. We verify that the agreement between TB and DFT bands is very satisfactory. Notice that, due to interlayer coupling, the type-II Dirac cones are doubled, as expected. Moreover, one of the cones crosses the Fermi energy for the three proposed stackings. This is an important feature for transport applications.
Notice that the split bands forming the Dirac cones are more separated in the AA case, whereas for the AB and AX there is a clear asymmetry in the splittings:
The flat bands are rather separated, but the bands with a larger slope are much closer in energy. An energy zoom allows us to distinguish two types of crossings, as depicted in Fig. \ref{FIG4} (d).
Notice that the bands have either a tic-tac-toe shape, with four Dirac points, as for AA and AB stackings, or a partial anticrossing, with only two bands crossings, for the 
AX. We denote this latter band crossing shape as "butterfly wings".

\section{Armchair nanoribbons}

\begin{table}[h]
\small
\caption{\label{tab:table1}%
Bilayer lattice parameters obtained from the DFT relaxed structures within vdW-DF2.
}\label{tablebilayer}
  \begin{tabular*}{0.48\textwidth}{@{\extracolsep{\fill}}ll}
    \hline
    Stackings & $d_{\perp}$  \\
    \hline
AA & 3.51 Å   \\
AB & 3.39 Å  \\
AX & 3.43 Å \\

    \hline
  \end{tabular*}
\end{table}

\begin{figure}[!h]
    \centering
    \includegraphics[width=8.8cm]{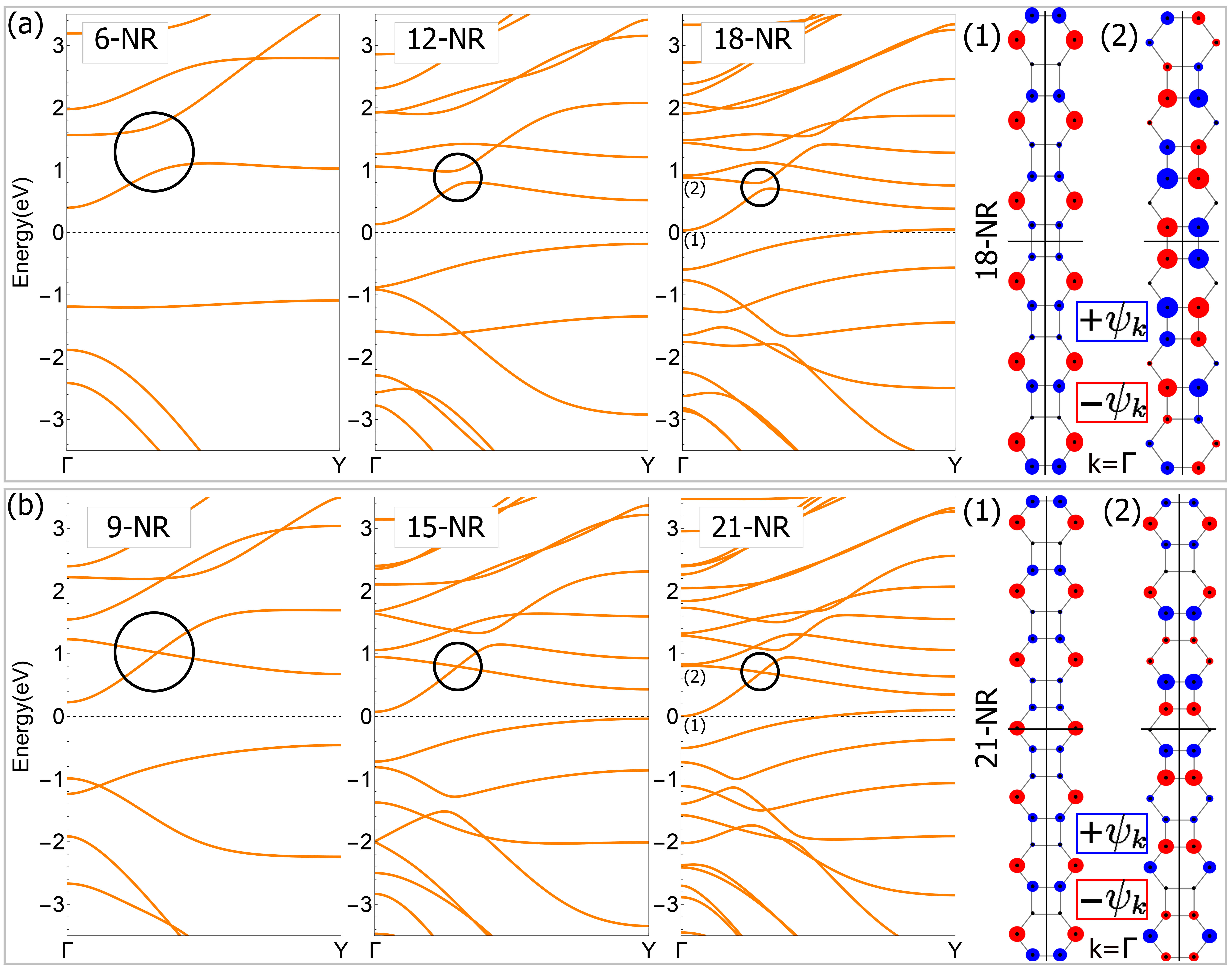}
     \caption{Electronic band structures of armchair nanoribbons with (a) even (6-NR, 12-NR, and 18-NR) and (b) odd (9-NR, 15-NR, and 21-NR) number of sites along the width. Wavefunction $\psi_{k=\Gamma}$ values and signs for the 18-NR and 21-NR, represented by the sizes and colors of circles within the unit cell, respectively, for bands (1) and (2). The TB parameters are $t_1$=-3.3 eV, $\beta=2.6$, and $\varepsilon=-1.3$ eV.}
\label{fig5}
\end{figure}

\begin{figure*}
    \centering
    \includegraphics[width=16cm, height=10cm]
    {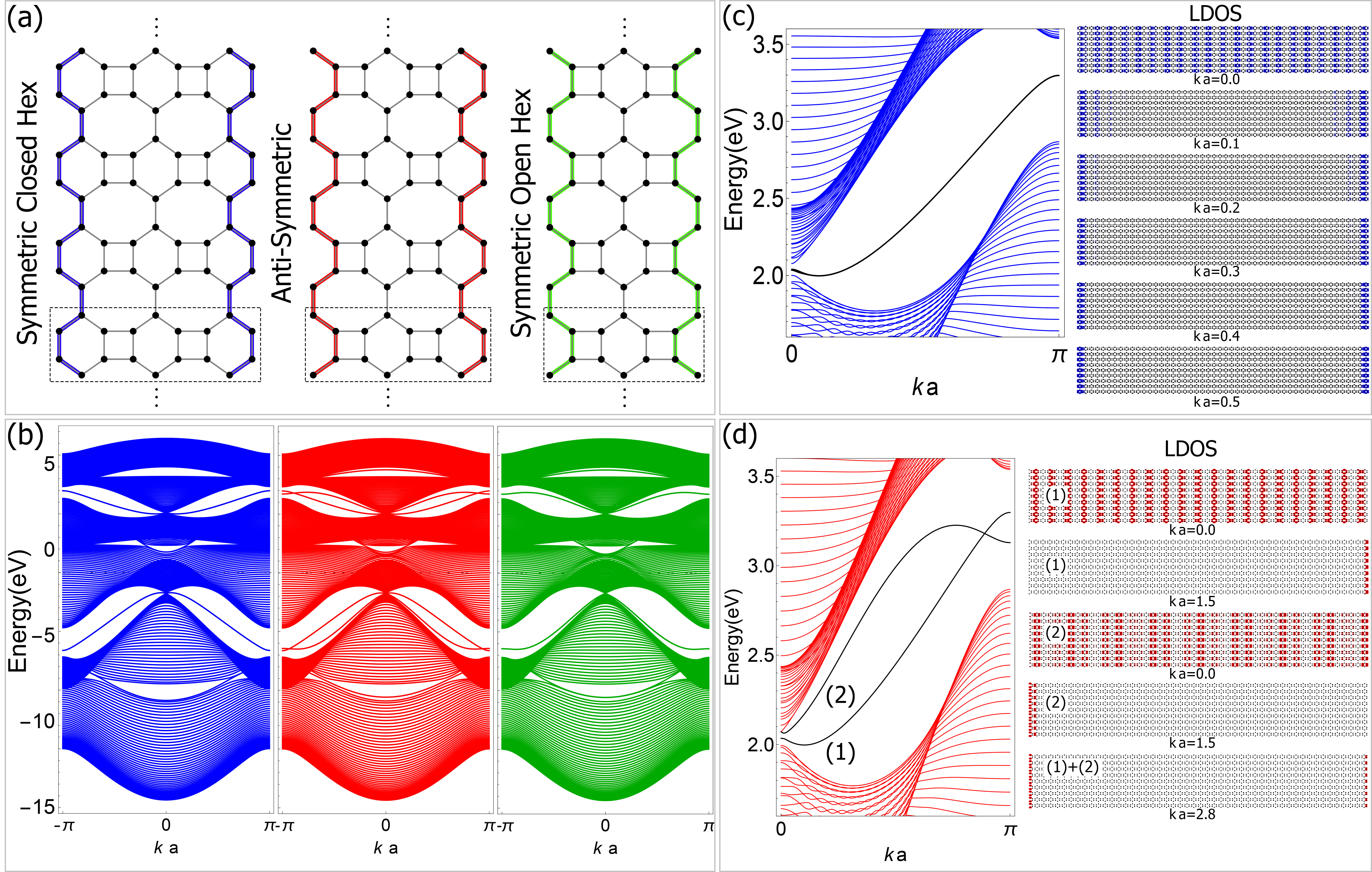}%v2(5)fig4
     \caption{(a) 9-NR Armchair nanoribbons geometries for three edge configurations in blue, red and green, respectively: symmetric closed hexagons (SCH), anti-symmetric (AS), and symmetric open hexagons (SOH). (b) Electronic bands  of 50-NRs for the three edge configurations ($\beta=2.6$ and $\varepsilon=0$). Zoom of the electronic bands of the (c) SCH and (d) AS 50-NRs and projected LDOS for different $ka$ values over the nanoribbon sites highlighted in black in panel (a). Parameters $t_1$=-3.3 eV, $\beta=2.6$, and $\varepsilon=-1.3$ eV.}\label{fig6}
\end{figure*}

In what follows, we perform biphenylene nanoribbon (NR) calculations to verify the efectiveness and convenience of the proposed TB parametrization in such quasi-1D  systems.
The atomic coordinates are obtained from the relaxed DFT (GGA-PBE) calculations of the monolayer, without performing further relaxations to take into account edge effects. We have verified by comparison to other DFT calculations \cite{fan2021biphenylene} that the edge effects are not important in such nanoribbons; in fact, our results are quite similar to the relaxed ones reported therein. 

We focus on the armchair edges, which are known to have semiconductor nature for smaller widths. 
An optimal parametrization 
has the same in-plane hopping $t_1=3.3$ eV used for the 2D system, and an adjusted on-site energy $\varepsilon$ and exponential parameter $\beta$; $\varepsilon=-1.3$ eV and $\beta=2.6$, respectively. These modifications are necessary to better describe 
confined states in quasi-1D periodic systems, in contrast to relaxed DFT nanoribbons from previous energy band results \cite{fan2021biphenylene}.

A type-II Dirac cone is also observed for nanoribbons with an odd number of hexagons %inside the unit cell 
across their width. However, for those with an even width, a gap opens, destroying the cone. This behavior is highlighted in Fig. \ref{fig5} (a) and (b) by black circles at the anticrossing or crossing regions, for even and odd ribbons, respectively. The projected wavefunctions
inside the unit cells at the $\Gamma$ point for the 18-NR and 21-NR
are depicted at the right side of the respective band structures in Fig. \ref{fig5}. The sign of the wavefunction is represented by red and blue colors, and their amplitudes by the disk sizes. 
Notice that the wavefunctions of the low-lying band at $\Gamma$, labeled (1), presents two mirror symmetries with respect to the horizontal ($M_h$) and vertical ($M_v$) planes marked with black lines, as well as inversion symmetry ($I$). However, the wavefunction of the the upper band, labeled (2), has fewer  different symmetries, which vary with the nanoribbon width. In the case shown in Fig. \ref{fig5}(a), the 18-NR, it is $M_h$; for Fig. \ref{fig5}(b), corresponding to 21-NR, it is $M_v$. The horizontal mirror symmetry is important for the confined wavefunctions; note that the probability density is symmetric with respect to the longitudinal axes of the ribbons. Therefore, if the wavefunctions  of bands (1) and (2) share this symmetry (for even widths), their bands anticross, whereas for odd ribbons, for which these bands do not share this reflection symmetry, the bands cross and the Dirac cone is preserved.  This even-odd behavior is identical to the parity change of the wavefunctions for successive states in textbook quantum wells with respect to its center. Here, the wavefunctions change differently with respect to the different mirror reflections; however, the important mirror symmetry for this even-odd behavior is the reflection with respect to the longitudinal axes. 
Similar symmetry arguments have successfully explained even-odd effects in other graphene-based nanoribbons \cite{erevis2020} and slabs of topological materials \cite{Araujo2018, ArroyoGascon2022}. 

The same parametrization was also employed for the study of wide ribbons. In Fig. \ref{fig6}(a) we present three edge configurations in the armchair features: symmetric closed hexagons (SCH), anti-symmetric (AS), and symmetric open hexagons (SOH), represented by blue, red, and green edges, respectively. Using the same color scheme, the electronic bands for each edge configuration are shown at Fig. \ref{fig6}(b) with parameters $t_1$=-3.3 eV, $\beta=2.6$, and $\varepsilon=-1.3$ eV. The energy states are on average the same for the three geometries considered. However, distinguishable isolated states located between the bulk energy bands, within the pseudogaps, are found for each of the three edge symmetries. 

A zoom in the two-fold degenerate SCH band structure (blue curves) is shown in Fig. \ref{fig6}(c). The LDOS is computed for this specific state between $ka=0$ and $ka=0.5$. At $ka=0$, we note that the charge density is spread over all the nanoribbon, because at this point the edge bands meet the bulk bands. A  change in the charge distribution happens, however, as shown for $ka=0.1$ up to $ka=0.5$, at the Brillouin  zone boundary.  Highly localized  states emerge at the nanoribbon edges for $ka\neq 0$. Comparing with the other symmetric edge nanoribbon SOH, such in-gap edge states emerging in the  electronic bands, are also degenerate as for the SCH case.  With respect to the AS nanoribbon, the twofold degeneracy is broken in such localized states, resulting in complementary edge states for the (1) and (2) bands, respectively, at $ka=1.5$ as shown in the LDOS of Fig. \ref{fig6}(d). At the band crossing ($ka=2.8$), the charge densities are localized in both edges, as seen in the (1)+(2) LDOS case. 
Recently, a two-hopping tight-binding parametrization has found via Zak phase calculations that such edge states are topological \cite{Wakabayashi2024}. Actually, by employing our exponential hopping parametrization we have verified that those edge states may change in number and energy, suggesting different Zak phases in the involved cases.

\section{Conclusions}

We have derived a robust tight-binding (TB) parametrization based on DFT calculations to describe the electronic structure of various biphenylene systems, including monolayers, bilayers, and nanoribbons of different widths. We confirm the presence of a type-II Dirac cone in both monolayer and bilayer geometries, in agreement with previous works. By means of DFT calcluations, we obtain relaxed structures that allows us to fit our TB  model, capturing successfully the electronic properties of biphenylene-based structures. The TB parametrization preserves essential symmetries and ensures the type-II nature of the Dirac cone by extending hopping up to third neighboring group distances ($>3$ Å).

We have introduced two novel bilayer symmetric stackings, AB and AX, not studied before. The bands are split due to the interlayer interaction, and one of the type-II Dirac cones is placed at the Fermi energy, suggesting new transport responses. THe novel AB and AX stackings   present similar total energies compared to previously studied AA bilayer biphenylene, which indicates that they are stable. 

Our combined approach correctly describes the type-II Dirac cone in both 2D and quasi-1D biphenylene structures, revealing crossings and anti-crossings in the Dirac cone for nanoribbons depending on their width, due to different symmetries in their corresponding wavefuntions. 
Additionally, for wider nanoribbons, our calculations have allowed the  identification of robust edge states localized at the edges of the unit cell. We expect that our findings motivate further theoretical and experimental work. Moreover, the present study is also applicable to other carbon-based systems with hybrid geometric symmetries, like penta-composites and others allotropes such graphenylene.

\section*{Acknowledgments}
The authors would like to thank the INCT de Nanomateriais de Carbono for providing support on the computational infrastructure. LLL thanks the CNPq scholarship. AL thanks the CNPq and FAPERJ under grants E-26/202.567/2019 and E-26/200.569/2023. LC and OAG acknowledge financial support from the Agencia Estatal de Investigación of Spain under grant PID2022-136285NB-C31, and from grant (MAD2D-CM)–(UCM5), Recovery, Transformation and Resilience Plan, funded by the European Union - NextGenerationEU. OAG acknowledges the support of grant PRE2019-088874 funded by MCIN/AEI/10.13039/501100011033 and by “ESF Investing in your future”. J.D. Correa is gratefully acknowledged for illuminating conversations. 
\\
\section*{Conflict of Interest Statement}
The authors declare that the research was conducted in the absence of any commercial or financial relationships that could be construed as a potential conflict of interest.

\section*{Author Contributions}

  AL and LC have conceived the idea of the present study. All the authors discussed the results and contributed to the final manuscript. AL and LC have contributed in the Writing – review and editing the manuscript, Conceptualization, Methodology, Formal Analysis, and Project administration. LL and OAG have contributed with the Conceptualization, Methodology, and in Writing – original and draft manuscript and editing.

%%%REFERENCES%%%
\bibliography{refs} 

\end{document}